\documentstyle[prl,psfig,aps,floats]{revtex}
\begin{document}
\draft
\lefthyphenmin=2
\righthyphenmin=3
\def\dsdpt{\mbox{$d\sigma^W/dp_T^W$}}
\def\etal{\it et al}
\newcommand{\HERW}{{\sc herwig}}

\title{\bf
Measurement of the ratio of differential cross sections
for  {\boldmath $W$} and 
{\boldmath $Z$} boson production as a function of transverse momentum
       in {\boldmath{\mbox{$p\bar p$}}}\ collisions at 
       {\boldmath{\mbox{$\sqrt{s}$ =\ 1.8\ TeV}}} }                                             
% LIST_OF_AUTHORS.TEX                 7/3/01             
%
\author{                                                                      
%% names begin here                                                           
V.M.~Abazov,$^{23}$                                                           
B.~Abbott,$^{58}$                                                             
A.~Abdesselam,$^{11}$                                                         
M.~Abolins,$^{51}$                                                            
V.~Abramov,$^{26}$                                                            
B.S.~Acharya,$^{17}$                                                          
D.L.~Adams,$^{60}$                                                            
M.~Adams,$^{38}$                                                              
S.N.~Ahmed,$^{21}$                                                            
G.D.~Alexeev,$^{23}$
% ...
\\                                                                            
\vskip 0.30cm                                                                 
\centerline{(D\O\ Collaboration)}                                             
\vskip 0.30cm                                                                 
}                                                                             
\address{                                                                     
\centerline{$^{1}$Universidad de Buenos Aires, Buenos Aires, Argentina}       
\centerline{$^{2}$LAFEX, Centro Brasileiro de Pesquisas F{\'\i}sicas,         
                  Rio de Janeiro, Brazil}                                     
\centerline{$^{3}$Universidade do Estado do Rio de Janeiro,                   
                  Rio de Janeiro, Brazil}                                     
\centerline{$^{4}$Institute of High Energy Physics, Beijing,                  
                  People's Republic of China}                                 
\centerline{$^{5}$Universidad de los Andes, Bogot\'{a}, Colombia}             
\centerline{$^{6}$Charles University, Center for Particle Physics,            
                  Prague, Czech Republic}                                     
\centerline{$^{7}$Institute of Physics, Academy of Sciences, Center           
                  for Particle Physics, Prague, Czech Republic}               
\centerline{$^{8}$Universidad San Francisco de Quito, Quito, Ecuador}         
\centerline{$^{9}$Institut des Sciences Nucl\'eaires, IN2P3-CNRS,             
                  Universite de Grenoble 1, Grenoble, France}                 
\centerline{$^{10}$CPPM, IN2P3-CNRS, Universit\'e de la M\'editerran\'ee,     
                  Marseille, France}                                          
\centerline{$^{11}$Laboratoire de l'Acc\'el\'erateur Lin\'eaire,              
                  IN2P3-CNRS, Orsay, France}                                  
\centerline{$^{12}$LPNHE, Universit\'es Paris VI and VII, IN2P3-CNRS,         
                  Paris, France}                                              
\centerline{$^{13}$DAPNIA/Service de Physique des Particules, CEA, Saclay,    
                  France}                                                     
\centerline{$^{14}$Universit{\"a}t Mainz, Institut f{\"u}r Physik,            
                  Mainz, Germany}                                             
\centerline{$^{15}$Panjab University, Chandigarh, India}                      
\centerline{$^{16}$Delhi University, Delhi, India}                            
\centerline{$^{17}$Tata Institute of Fundamental Research, Mumbai, India}     
\centerline{$^{18}$Seoul National University, Seoul, Korea}                   
\centerline{$^{19}$CINVESTAV, Mexico City, Mexico}                            
\centerline{$^{20}$FOM-Institute NIKHEF and University of                     
                  Amsterdam/NIKHEF, Amsterdam, The Netherlands}               
\centerline{$^{21}$University of Nijmegen/NIKHEF, Nijmegen, The               
                  Netherlands}                                                
\centerline{$^{22}$Institute of Nuclear Physics, Krak\'ow, Poland}            
\centerline{$^{23}$Joint Institute for Nuclear Research, Dubna, Russia}       
\centerline{$^{24}$Institute for Theoretical and Experimental Physics,        
                   Moscow, Russia}                                            
\centerline{$^{25}$Moscow State University, Moscow, Russia}                   
\centerline{$^{26}$Institute for High Energy Physics, Protvino, Russia}       
\centerline{$^{27}$Lancaster University, Lancaster, United Kingdom}           
\centerline{$^{28}$Imperial College, London, United Kingdom}                  
\centerline{$^{29}$University of Arizona, Tucson, Arizona 85721}              
\centerline{$^{30}$Lawrence Berkeley National Laboratory and University of    
                  California, Berkeley, California 94720}                     
\centerline{$^{31}$University of California, Davis, California 95616}         
\centerline{$^{32}$California State University, Fresno, California 93740}     
\centerline{$^{33}$University of California, Irvine, California 92697}        
\centerline{$^{34}$University of California, Riverside, California 92521}     
\centerline{$^{35}$Florida State University, Tallahassee, Florida 32306}      
\centerline{$^{36}$University of Hawaii, Honolulu, Hawaii 96822}              
\centerline{$^{37}$Fermi National Accelerator Laboratory, Batavia,            
                   Illinois 60510}                                            
\centerline{$^{38}$University of Illinois at Chicago, Chicago,                
                   Illinois 60607}                                            
\centerline{$^{39}$Northern Illinois University, DeKalb, Illinois 60115}      
\centerline{$^{40}$Northwestern University, Evanston, Illinois 60208}         
\centerline{$^{41}$Indiana University, Bloomington, Indiana 47405}            
\centerline{$^{42}$University of Notre Dame, Notre Dame, Indiana 46556}       
\centerline{$^{43}$Iowa State University, Ames, Iowa 50011}                   
\centerline{$^{44}$University of Kansas, Lawrence, Kansas 66045}              
\centerline{$^{45}$Kansas State University, Manhattan, Kansas 66506}          
\centerline{$^{46}$Louisiana Tech University, Ruston, Louisiana 71272}        
\centerline{$^{47}$University of Maryland, College Park, Maryland 20742}      
\centerline{$^{48}$Boston University, Boston, Massachusetts 02215}            
\centerline{$^{49}$Northeastern University, Boston, Massachusetts 02115}      
\centerline{$^{50}$University of Michigan, Ann Arbor, Michigan 48109}         
\centerline{$^{51}$Michigan State University, East Lansing, Michigan 48824}   
\centerline{$^{52}$University of Nebraska, Lincoln, Nebraska 68588}           
\centerline{$^{53}$Columbia University, New York, New York 10027}             
\centerline{$^{54}$University of Rochester, Rochester, New York 14627}        
\centerline{$^{55}$State University of New York, Stony Brook,                 
                   New York 11794}                                            
\centerline{$^{56}$Brookhaven National Laboratory, Upton, New York 11973}     
\centerline{$^{57}$Langston University, Langston, Oklahoma 73050}             
\centerline{$^{58}$University of Oklahoma, Norman, Oklahoma 73019}            
\centerline{$^{59}$Brown University, Providence, Rhode Island 02912}          
\centerline{$^{60}$University of Texas, Arlington, Texas 76019}               
\centerline{$^{61}$Texas A\&M University, College Station, Texas 77843}       
\centerline{$^{62}$Rice University, Houston, Texas 77005}                     
\centerline{$^{63}$University of Virginia, Charlottesville, Virginia 22901}   
\centerline{$^{64}$University of Washington, Seattle, Washington 98195}       
}                                                                             
%end                                                                          

\maketitle                          
\begin{abstract}

We report on a 
measurement of the ratio of the differential cross sections 
for $W$ and $Z$ boson production as a function of transverse momentum 
in proton-antiproton collisions at {\mbox{$\sqrt{s}$ =\ 1.8\ TeV}}.
This measurement uses data recorded by the D\O\ detector at 
the Fermilab Tevatron in 1994--1995. It represents the first investigation
of a proposal that ratios between $W$ and $Z$ observables
can be calculated reliably using perturbative QCD, even when the individual
observables are not.
Using the ratio of differential cross sections
reduces both experimental and theoretical uncertainties,
and can therefore provide smaller overall uncertainties in the measured 
mass and width of the $W$ boson than current  methods used 
at hadron colliders.
\end{abstract}
\pacs{PACS numbers 12.35.Qk,14.70.Fm,12.38.Qk}

\section{Introduction}
The D\O\ Collaboration 
has recently published~\cite{wpt,zpt} measurements of 
differential cross sections for $W$ and $Z$ boson production  
as a function of transverse momentum ($p_T$). Both measurements are in good
agreement with combined resummed and perturbative QCD models, such
as those in Refs.~\cite{ak,ly,ev}. For the analyses of data taken
during 1992--1996 (Fermilab Tevatron Run 1), we have 
used the resummed calculation of Ref.~\cite{ly} fitted to our observed
$Z \to e^{+} e^{-}$ differential cross section 
to extract the non-perturbative phenomenological parameters of the
theory. The resummed calculation was then used to predict 
$W$ boson observables such as the
electron and neutrino transverse momenta and as
input to a Monte
Carlo model of $W$ boson production and decay, which we used to 
extract the mass~\cite{wmass} and
production cross section~\cite{xsections} of the $W$ boson.

Ref.~\cite{gk} proposes an alternative method of predicting $W$ boson
observables from measured $Z$ boson quantities. This is based on 
the theoretical
ratio of the $W$ to $Z$ boson differential cross sections with respect
to variables that have been scaled by their corresponding vector boson masses.
Because production properties of $W$ and $Z$ bosons are very similar, 
the large radiative corrections that affect the individual distributions 
cancel in the ratio. 
The ratio can therefore be calculated reliably
 using perturbative QCD (pQCD), 
with no need for resummation, even at
small values of the transverse momenta of the vector bosons, for which 
the radiative corrections factorize from the hard process and 
therefore cancel in the ratio. The theoretical uncertainties 
stemming from the perturbative expansion are consequently
well understood, and are smallest at very low $p_T$. 

The basic proposal of Ref.~\cite{gk}
is to use pQCD calculations and the measured
$Z$ boson observables to extract the $W$ boson observables. 
Compared to the standard method used previously 
to extract $W$ boson observables, 
the present method reduces
both theoretical and experimental systematic uncertainties. However,
it introduces a statistical contribution to the uncertainty
from the number of events in the $Z$ boson candidate sample.
Such a trade-off will eventually
result in smaller overall uncertainties, especially when used
with the high statistics samples expected from Run 2 of the Tevatron.

Corroborating 
the agreement of the pQCD calculation with data
is vital if the new procedure
is to be used to improve the measurement of the $W$ boson mass in future
collider runs.
In this letter, we will check the validity of the method
using the measured  differential cross 
sections for $W$ and $Z$ boson production 
as a function of transverse momentum. 
Both distributions were 
measured at the Tevatron~\cite{wpt,zpt,zptcdf},
where the systematic uncertainty on the $p_T^W$ 
at lowest transverse momentum is
four times larger than the corresponding uncertainty in $p_T^Z$.
The uncertainty in $p_T^Z$  is dominated by 
statistics. Hence, once large samples of $Z$ boson events become available,
it is expected that, if theoretical
uncertainties can be kept small, 
using the pQCD prediction and the well-measured $p_T^Z$
distribution to predict the $p_T^W$ distribution should
lead to smaller overall uncertainties 
on the measured mass  and width of the $W$ boson, 
relative to current methods used at hadron colliders.

The main 
difference between the production properties of the $W$ and the $Z$ bosons 
arises from the difference in their masses. 
We will therefore introduce variables that are scaled by
the corresponding vector boson mass $M_V$.
The ratio of differential cross sections for the scaled $W$  and
$Z$ boson transverse momenta ($p_T^W/M_W$ and $p_T^Z/M_Z$) is defined as
\begin{equation}
R_{p_T}=\left[\frac{d\sigma^W}{d(p_T^W/M_W)}\right] {\Big /}
\left[ \frac{d\sigma^Z}{d(p_T^Z/M_Z)} \right] ,
\label{eq:ratiodef}
\end{equation}
where $d\sigma^V/dp_T^V$ is the standard differential cross section for
vector boson production $\sigma(p\bar p \to V + X)$ as a function of 
transverse momentum $p_T^V$.
Equation~\ref{eq:ratiodef} can be used to predict the differential
cross section for $W$ bosons with respect to the non-scaled 
transverse momentum~\cite{gk}:
\begin{equation}
\left. \frac{d\sigma^W}{dp_T^W}  \right|_{\rm predicted}=
\frac{M_Z}{M_W} \times R_{p_T} \times  
\left. \frac{d\sigma^Z}{dp_T^Z}  \right|_{\rm measured}^{p_T^Z=\frac{M_Z}{M_W} p_T^W} ,
\label{eq:wptdef}
\end{equation}
where $R_{p_T}$ is calculated using pQCD.
In this paper, we present the first measurement of $R_{p_T}$, and compare it
to the calculation of Ref.~\cite{gk}. 
For completeness, we repeat the exercise presented in  Ref.~\cite{gk} and
use our measured differential $Z$ boson cross section in
Eq.~\ref{eq:wptdef}, and $R_{p_T}$ from
Ref.~\cite{gk}, to obtain the differential 
$W$ boson cross section and compare it to our 
published result~\cite{wpt}.

\section{Data Selection}
We keep modifications to the published D\O\
analyses~\cite{wpt,zpt} to a minimum, but, at the same time, we try to
cancel as many experimental 
uncertainties as possible in measuring $R_{p_T}$. The 
uncertainty in the integrated luminosity of the data samples ($4.3\%$) is the
dominant uncertainty in the individual cross sections.
It cancels completely when taking the ratio, as long as the same
data sets are used to select the final $W$ and $Z$ boson candidate samples.
In this analysis, we keep the event selections and 
corrections for background, efficiency, acceptance, and detector resolutions
identical to those in the published results~\cite{wpt,zpt}, 
but require total overlap in the data-taking runs for the 
$W$ and $Z$ boson event samples. In addition, we exclude events at collision
times with
large beam losses from the Main Ring accelerator~\cite{xsections}. 
These beam losses can create significant
energy deposits in the calorimeter that produce events with large false 
transverse momentum imbalance that could pass our $W$ boson selection criteria.
Due to these additional requirements, 
the $Z$ boson sample was reduced from 6407 to 4881 events.
About half of the events were lost due to tightening of 
beam quality conditions, and half because the 
$W$ trigger was not available or was prescaled.
The $W$ sample was reduced from 50488 to 50264 events when we removed
runs for which the $W$ trigger was prescaled.
The final 
integrated luminosity for both samples is $(84.5 \pm 3.6)\rm pb^{-1}$. 

We have investigated whether 
additional sources of error could be cancelled in the
ratio. There are four sources of systematic error that contribute 
to the $W$ and $Z$ boson cross sections.
These arise from uncertainties in the background estimate, the event selection 
efficiency, and the unfolding procedure used to correct for acceptance and 
detector resolution.

The dominant sources of background in both the
$W$ and the $Z$ boson analyses are from multijet and photon-jet events, where
the jets pass our electron identification criteria. In the case of the $W,$
a large imbalance in the transverse energy has to arise to mimic the presence
of a neutrino. The way multijet or photon-jet events 
mimic $W$ or $Z$ boson events is quite 
different, and the methods used to estimate background are 
independent. We therefore cannot cancel any contribution to the error in the
ratio arising from background estimates.

Acceptance and unfolding corrections are applied 
using a parameterized
Monte Carlo~\cite{wmass}. The main contribution to the error is from the
modeling of the detector. 
For the $W$ analysis, we rely completely on the measurement of the 
energy of the recoiling hadrons, whereas for 
the $Z$ boson measurement we use the electromagnetic energy deposited by
the electrons. We therefore do not benefit from cancellation of errors
in the acceptance/unfolding procedure.

The uncertainty in the 
efficiency has contributions from the trigger and offline
electron identification. The Level 0 trigger, 
which requires the detection
of an inelastic collision via simultaneous hits in the forward and
backward Level 0 scintillation detectors~\cite{d0detector}, 
is common for $W$ and $Z$ boson events.
The uncertainty in this trigger therefore cancels completely 
in the ratio. However, its contribution to the error in the
efficiency is negligible (0.5\% out of a total of 3.5\%). 

Although the triggers and the offline electron identification criteria 
used in the $W$ and $Z$ boson analyses are different, 
the main contribution (3\%) to the error in the efficiency  comes from
a common source, the so-called $u_{||}$ efficiency~\cite{wmass}.
This inefficiency arises
when the energy flow close to the electron increases as 
recoiling hadrons approach 
the electron. It is therefore a topological effect produced
by the proximity of the electron to the jet, and has the largest effect
at a boson transverse momentum of about $20\;\rm GeV$~\cite{zpt}.
The $u_{||}$ efficiency 
is calculated on an electron-by-electron
basis using the parameterized Monte Carlo.
The error in the $u_{||}$ efficiency is estimated from 
$W$ and $Z$ boson events, generated in \HERW~\cite{herwig}, and 
overlaid with data taken from randomly selected
{\mbox{$p\bar p$}}\ collisions. Because this
inefficiency depends on the proximity of electrons to jets,
it is difficult to estimate
how much of the uncertainty in the
$u_{||}$ efficiency cancels in the ratio. 
To determine if further investigation of any possible
cancellation of the uncertainty in $u_{||}$ efficiency was warranted, 
we estimated the effect on $R_{p_{T}}$
of a complete cancellation of the contribution from the uncertainty in
$u_{||}$ efficiency. This produced a maximum
reduction of uncertainty in $R_{p_{T}}$ of less than 5\%. 
We therefore concluded that no 
cancellations beyond the uncertainty in the luminosity would improve
significantly the measurement of $R_{p_{T}}$.

\section{\boldmath Scaled $W$ and $Z$ Boson Cross Sections}

Equation~\ref{eq:ratiodef} can be written
%
%\begin{equation}
\[ R_{p_T}^{\rm th}=\left(\frac{d\sigma^W}{dp_T^W}\right) {\Big /}
\left( \frac{d\sigma^Z}{d(p_T^Z \times M_W/M_Z)} \right), \]
%\label{eq:ratiodef2} 
%\end{equation}
\noindent
where we use the mass ratio from the Review of Particle Physics~\cite{pdb} 
%\begin{equation}
\[ \frac{M_W}{M_Z}=0.8820\pm0.0005 . \]
%\label{eq:massratio}
%\end{equation}
\noindent
In order to measure the scaled distributions without changing the
$p_T$-binning of both the $W$ and $Z$ 
boson analyses, we keep the $W$ bin boundaries
($\delta_i$) identical to the ones in our published work, but because 
we require the same bin widths in the scaled variables $p_T^W/M_W$
and $p_T^Z/M_Z$, we set the bin boundaries in the differential $Z$ boson 
cross section to $\delta_i/0.8820,$
and recompute the differential $Z$ boson cross section accordingly.

Table~\ref{tab:d0data} shows the modified results for the
$W$ and $Z$ boson cross sections, with the statistical
and systematic contributions to the uncertainties shown separately. 
It is clear that
the error in the ratio is dominated by the systematic uncertainty
in the $W$ cross section.

%\clearpage
\section{\boldmath Measurement of $R_{p_T}$}

\begin{figure}[h]
\centerline{\psfig{figure=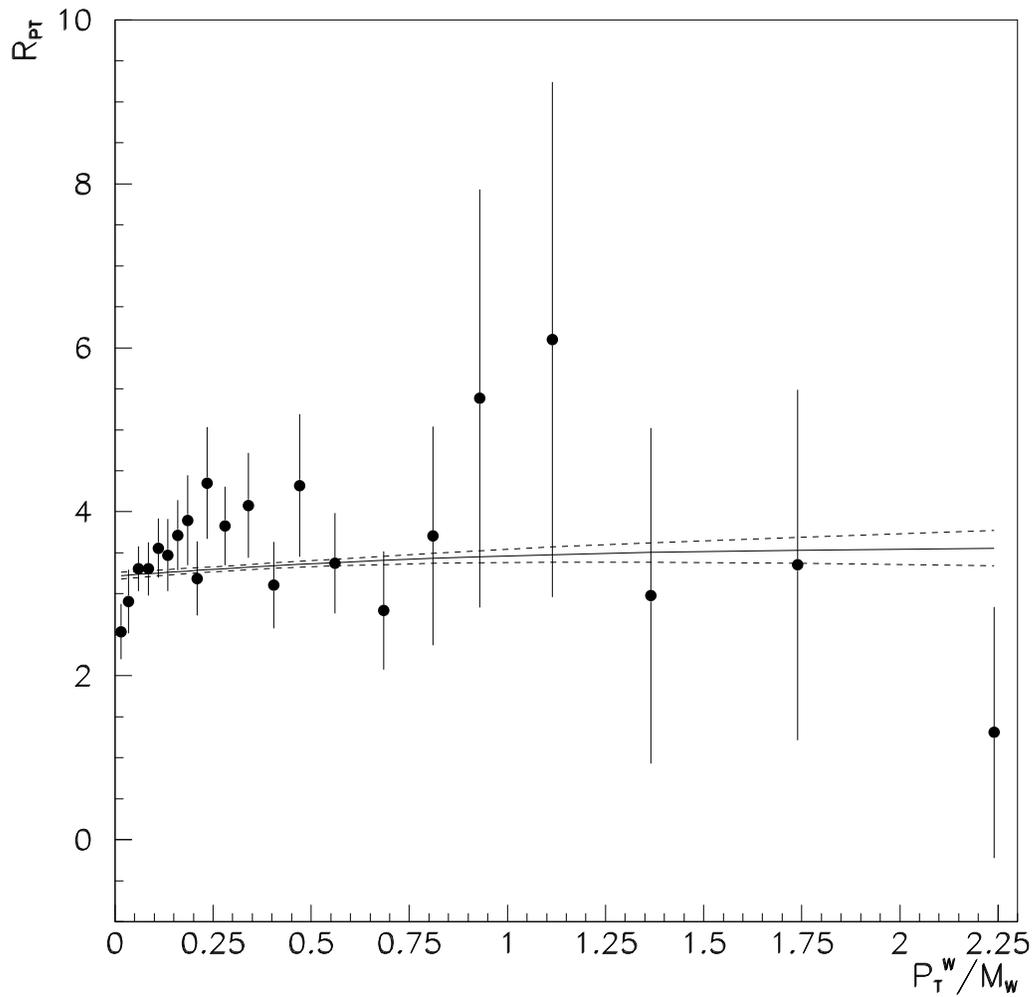,width=15cm}}
\caption{Ratio of scaled differential cross sections 
$R_{p_T}$ for $W$ and $Z$ production.
The solid line is the order $\alpha_S^2$ theoretical prediction of
Ref.~\protect\cite{gk}, and  the dotted lines are the one standard
deviation  uncertainties
due to Monte Carlo integration.
The error in the luminosity cancels completely in the ratio of the
measured cross sections.}
\label{fig:rpt}
\end{figure}

Based on the measured $W$ and $Z$ boson differential cross sections listed in 
Table~\ref{tab:d0data}, we extract the ratio of scaled cross sections
as a function of $p_T$:
%\begin{equation}
\[ R_{p_T}^{\rm exp}=\left[ \left(\frac{d\sigma^W}{dp_T^W}\right) {\Big /}
\left( \frac{d\sigma^Z}{dp_T^Z} \right)\right] \times
\frac{M_W}{M_Z}\times \frac{B(Z\to ee)}{B(W\to e\nu)} . \]
%\label{eq:dataratio}
%\end{equation}
It should be recognized 
that the prediction for $R_{p_T}$~\cite{gk} was calculated for the
ratio of the scaled $W$ and $Z$ boson differential cross sections
$d\sigma^V/dp_T^V$, but we
measure the differential cross sections multiplied by
their branching fractions to
electrons $(d\sigma^V/dp_T^V)\times B(V \to e)$. We therefore
must multiply our measurement by the ratio of the
$Z$ to $W$ boson branching fractions. Because the measurement of the
$W$ boson branching fraction from the Tevatron is obtained precisely
from the ratio of $W$ to $Z$ production cross sections~\cite{xsections},
we use the result from the LEP Electroweak Working Group~\cite{lep2} 
for the $W$ branching fraction,
to avoid a circularity problem. We take the value for the $Z$ branching 
fraction from the Review of Particle Physics~\cite{pdb}.
%\begin{equation}
\[ B(W\to e\nu)=0.1073 \pm 0.0025  \]
%\end{equation}
%\begin{equation}
\[ B(Z\to ee)=0.033632 \pm 0.000059 . \]
%\end{equation}
The result is shown in Fig.~\ref{fig:rpt}, and summarized in 
Table~\ref{tab:ratio}. 
The data are plotted at the value of
$p_T$ for which the theoretical prediction for $R_{p_T}$ is equal to its
average in the bin, following the prescription of Ref.~\cite{xbins}. 
We observe that the measured $R_{p_T}$
agrees with the pQCD
prediction~\cite{gk}: the $\chi ^2$ for the comparison
between data and theory is $18.3$ for 21 degrees of freedom (63\%
probability).
If we only consider the results in the first 12 bins,
the $\chi ^2$ is $12.8$ for 11 degrees of freedom, which corresponds to
a probability of 31\%. 

We should mention that, at this time, 
the only uncertainty included in the theoretical prediction is the one arising
from Monte Carlo integration.
Additional uncertainties must be considered to
determine whether the agreement between data and theory can be improved, 
in particular, if
$R_{p_T}$ should be calculated to higher orders, or whether non-perturbative 
effects are playing a role at lowest $p_T$. 
Once the theoretical uncertainties are improved, this would provide the
means for estimating the integrated luminosity at which
the ratio method will provide a superior measurement of the $W$ boson mass.

\section{\boldmath Extraction of $\dsdpt$}
\begin{figure}[h]
\centerline{\psfig{figure=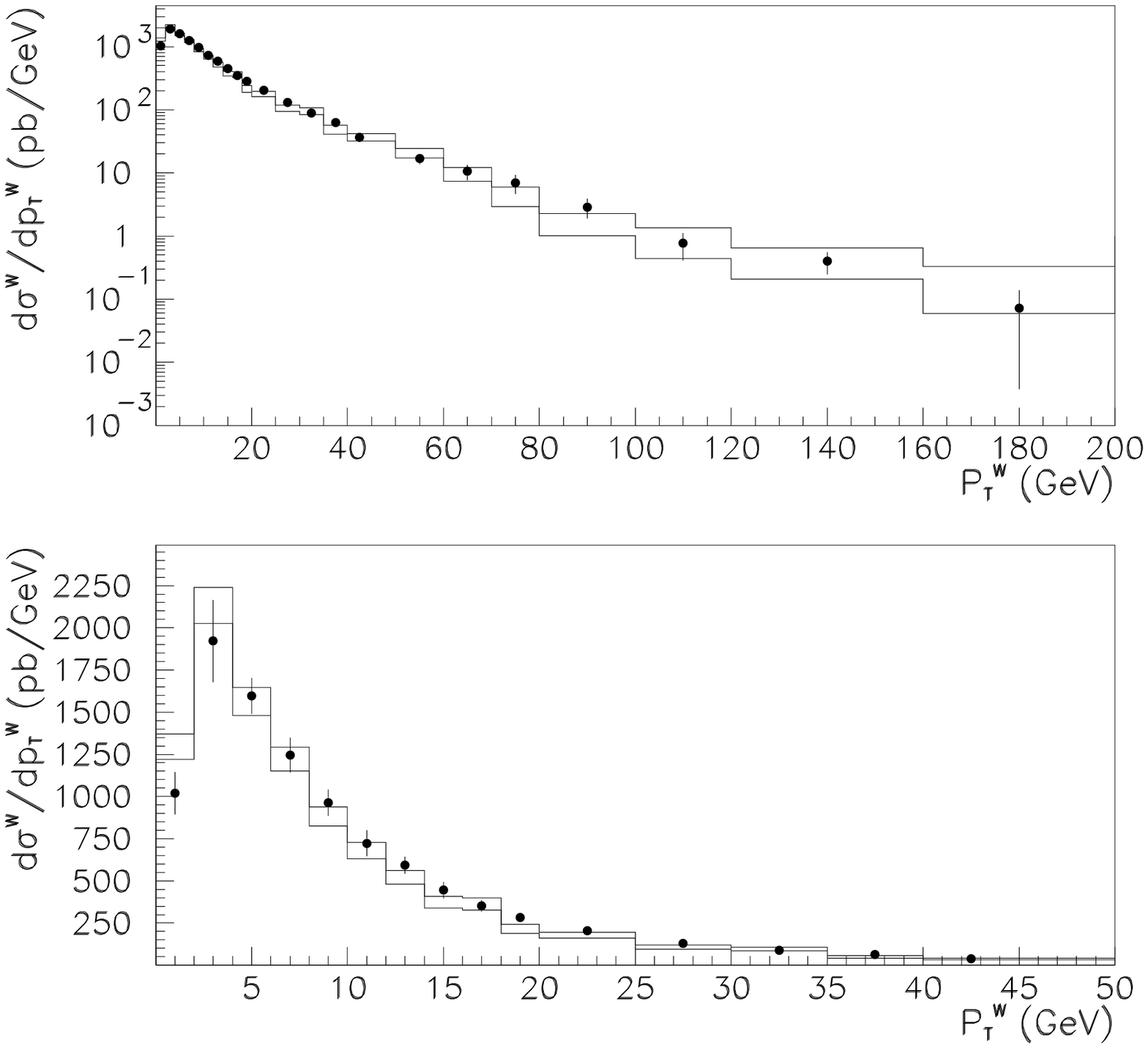,width=15cm}}
\caption{Differential cross section for $W$ boson production as a function
of $p_T^W$ shown for the entire $p_T^W$ range (upper plot) and the low
$p_T^W$ region (lower plot). 
The points are the D\O\ data; the error bars do not include
the $4.3\%$ error in the luminosity. The histograms represent the 
upper and lower 68\%
confidence level limits of the prediction~\protect\cite{gk} 
obtained from the ratio method.}
\label{fig:wpt}
\end{figure}

Based on Eq.~\ref{eq:wptdef}, we 
use the calculated $R_{p_T}$ in Ref.~\cite{gk}, 
together with the measured 
$d\sigma^Z/dp_T^Z$, to predict the $W$ boson transverse
momentum spectrum, and
compare it with our previously measured $d\sigma^W/dp_T^W$~\cite{wpt}. 
This is shown in Fig.~\ref{fig:wpt}, and is an update of the
result given in Ref.~\cite{gk} using our final data samples.
For simplicity, we use the measured $p_T^Z$ distribution from 
Table~\ref{tab:d0data}. A better prediction for $p_T^W$ can
be obtained from the combination of our published $p_T^Z$~\cite{zpt}
and the corresponding measurement from CDF~\cite{zptcdf}.
Fig.~\ref{fig:wpt} shows the measured differential cross section
plotted at the center of the bin. 
The upper and lower 68\% confidence level 
limits for the prediction are plotted as histograms.
The extracted transverse momentum distribution agrees well with the
measurement: the Kolmogorov-Smirnov probability~\cite{kstest} 
$\kappa$ is equal to 0.987.

\section{Conclusions}
We have measured the ratio of scaled differential cross sections $R_{p_T}$
for $W$ and $Z$ boson production, and compared it to a purely
pQCD prediction. 
We observe good agreement between data and theory
over the entire $p_T$ spectrum. For completeness, we have used
the theoretical prediction
for $R_{p_T}$, together with our measurement of the differential $Z$ boson 
production cross section, to extract the differential cross section
for $W$ production. As expected,
this prediction agrees with our published result.
From this first study of the method of Ref.~\cite{gk} for
predicting $W$ boson properties, we conclude that, 
once the high statistics samples of $Z$ boson events expected from 
Run 2 at the Tevatron become available, this new approach 
should lead to smaller overall uncertainties 
on the measured mass 
and width of the $W$ boson, 
compared to current methods used at hadron colliders.

\section*{Acknowledgments}
\label{sec:ack}
%
% =========> Choose the first acknowledgments paragraph <==========
% =========> if you are writing a physics paper. <==================
%
We thank Stephane Keller for interesting discussions
leading to this measurement, and Walter Giele for
providing the theoretical prediction and assistance with technical
details to complete this study.
% Acknowledgment_paragraph.tex
%
We also thank the staffs at Fermilab and collaborating institutions, 
and acknowledge support from the 
Department of Energy and National Science Foundation (USA),  
Commissariat  \` a L'Energie Atomique and 
CNRS/Institut National de Physique Nucl\'eaire et 
de Physique des Particules (France), 
Ministry for Science and Technology and Ministry for Atomic 
   Energy (Russia),
CAPES and CNPq (Brazil),
Departments of Atomic Energy and Science and Education (India),
Colciencias (Colombia),
CONACyT (Mexico),
Ministry of Education and KOSEF (Korea),
CONICET and UBACyT (Argentina),
The Foundation for Fundamental Research on Matter (The Netherlands),
PPARC (United Kingdom),
Ministry of Education (Czech Republic),
and the A.P.~Sloan Foundation.

%%%%%

\clearpage

\begin{table}[t]
\caption{Summary of the measured $W$ 
and $Z$ boson differential production cross sections as a function
of transverse momentum used to calculate the ratio. 
The error in the ratio is dominated by the systematic error
in the $W$ cross section.}
\begin{center}
\begin{tabular}{||c|c|c|c|c|c|c||}
$p_T$ Bin & {\mbox{$\frac{d\sigma(W\to e \nu)}{dp_T^{W}}$}}&
Stat Error & Syst Error & 
{\mbox{$\frac{d\sigma(Z\to e^+ e^-)}{dp_T^Z}$}}& Stat Error
& Syst Error\\
\scriptsize{(GeV)} & \scriptsize{(pb/GeV)} & \scriptsize{(pb/GeV)} &
\scriptsize{(pb/GeV)} & \scriptsize{(pb/GeV)}
& \scriptsize{(pb/GeV)} & \scriptsize{(pb/GeV)} \\
\hline \hline %$
 0--2 &  109.48 &  4.61   & 12.35 &      11.94 & 0.53 &  0.35 \\ \hline 
 2--4 &  206.21 & 6.85   & 24.64  &      19.63 & 0.65 &  0.57 \\ \hline
 4--6 &  171.32 & 5.65   & 9.29 &      14.34 & 0.53 & 0.44 \\ \hline
 6--8 & 133.60 &  4.65   & 9.46 &      11.19 & 0.48 &  0.36 \\ \hline
 8--10 & 103.48 &  4.04   & 6.95  &      8.05 & 0.41 &   0.27 \\ \hline
10--12 & 77.46 &  3.46   & 7.25 &     6.18 &    0.37 &  0.21 \\ \hline 
12--14 &  63.58 & 3.20   & 4.16  &    4.74 &   0.33 &   0.15 \\ \hline  
14--16 &   47.77&  2.77  & 4.29 &      3.39 &   0.28 &  0.11 \\ \hline
16--18 &  37.67 &  2.42     & 2.73 &       3.27 & 0.28 &   0.17 \\ \hline
18--20 & 30.50 &    2.20   & 1.74 &       1.94 & 0.22 &  0.11 \\ \hline
20--25 & 22.02 &    1.23   & 1.22 &      1.59&  0.12 &   0.08  \\ \hline
25--30 & 13.94&    0.93   &   1.07 &     0.946 & 0.097 &  0.051 \\ \hline
30--35 &  9.51 &   0.73   &  0.84 &     0.848 &  0.092 &   0.043 \\ \hline
35--40 &  6.79&    0.63   &  0.51&     0.435&   0.066 &  0.022 \\ \hline
40--50 & 3.96 &    0.37   &  0.31 &     0.325 & 0.040 &   0.016 \\ \hline
50--60 & 1.82 &   0.25   &  0.25 &     0.180 &  0.029 &  0.009  \\ \hline
60--70 & 1.14 &    0.20   &  0.23 &    0.0848 & 0.0197 &   0.0045 \\ \hline
70--80 &  0.749 &   0.178  &   0.170   &     0.0385 & 0.0129 & 0.0020 \\ \hline
80--100 &  0.310&   0.059 & 0.088 &     0.0141 &  0.0054 &   0.0008 \\ \hline
100--120 &  0.0822&  0.0287& 0.0255 & 0.00764& 0.00383 &  0.00032 \\ \hline
120--160 &  0.0433& 0.0119& 0.0118& 0.00358&  0.00180 &  0.00018 \\ \hline
160--200 &  0.00769& 0.00545 & 0.00482& 0.00163& 0.00111 &  0.00010 \\ 
\end{tabular}
\end{center}
\label{tab:d0data}
\end{table}

\clearpage

\begin{table}[t]
\caption{Measured $R_{p_T}$. The uncertainty in the luminosity
for the $W$ and $Z$ samples cancels completely when taking the
ratio. }
\begin{center}
\begin{tabular}{ ||c|c|c|c||}
$p_T$ Bin \scriptsize{(GeV)} & $p_T$ \scriptsize{(GeV)} & $R_{p_T}$ & 
Total Error \\ 
\hline \hline %$
 0--2 & 1.21 &  2.538  &   0.339 \\ \hline 
 2--4 &  2.81 & 2.908  &   0.388  \\ \hline
 4--6 &  4.83 &  3.306 &     0.275  \\ \hline
 6--8 &  6.84 &  3.305 &     0.324 \\ \hline
 8--10 & 8.85 &   3.557 &     0.361 \\ \hline
10--12 & 10.86 &   3.471 &     0.439\\ \hline 
12--14 & 12.87 &   3.714 &     0.426 \\ \hline  
14--16 & 14.88 &   3.895 &     0.549 \\ \hline
16--18 & 16.89 &  3.187 &     0.449 \\ \hline
18--20 & 18.90 &   4.351 &     0.681\\ \hline
20--25 & 22.52 & 3.829 &    0.478 \\ \hline
25--30 & 27.34 & 4.078 &     0.638 \\ \hline
30--35 & 32.57 & 3.104 &     0.528 \\ \hline
35--40 & 37.89 &  4.320 &     0.871 \\ \hline
40--50 & 45.03 &  3.373 &     0.613  \\ \hline
50--60 & 55.09 & 2.796 &     0.724\\ \hline
60--70 & 65.14 &  3.707 &      1.334\\ \hline
70--80 & 74.79 &  5.384 &      2.551 \\ \hline
80--100 & 89.67 &  6.100 &      3.141 \\ \hline
100--120 & 109.77 & 2.976 &      2.047 \\ \hline
120--160 & 139.93 & 3.352 &      2.138  \\ \hline
160--200 & 180.14 & 1.309 &      1.529 \\ 
\end{tabular}
\end{center}
\label{tab:ratio}
\end{table}
\clearpage

\clearpage

\clearpage

\begin{thebibliography}{}
%LIST_OF_VISITOR_ADDRESSES.TEX
\bibitem[*]{lehner}
Visitor from University of Zurich, Zurich, Switzerland.
%
\vskip 0.25cm

\bibitem{wpt} V.\ M.\ Abazov {\etal.}, (D\O\ Collaboration), 
to be published in  Phys.\ Lett.\ B,
     FERMILAB-Pub-00/268-E, hep-ex/0010026.
\bibitem{zpt} B.\ Abbott {\etal.}, (D\O\ Collaboration), 
     Phys.\ Rev.\ D 61, 032004 (2000), and Phys.\ Rev.\ Lett.\
84, 2792 (2000).
\bibitem{ak} P.\ Arnold and R.\ Kauffman, 
Nucl.\ Phys.\ B349, 381 (1991).
%
\bibitem{ly} G.\ A.\ Ladinsky and C.\ P.\ Yuan,
Phys.\ Rev.\ D 50, 4239 (1994).
%
\bibitem{ev} R.~K.~Ellis and S.~Veseli,
Nucl.\ Phys.\  B511, 649 (1998).
%
\bibitem{wmass} S.\ Abachi {\etal.}, (D\O\ Collaboration), 
Phys.\ Rev.\ Lett.\ 77, 3309 (1996); 
{\sl ibid.} 80, 3008 (1998);
{\sl ibid.} 84, 222 (2000);
Phys.\ Rev.\ D 58, 12002 (1998); 
{\sl ibid.} D 58, 092003 (1998); 
{\sl ibid.} D 62, 092006 (2000).
%
\bibitem{xsections}  B.\ Abbott {\etal.}, (D\O\ Collaboration),
Phys.\ Rev.\ D 61, 072001 (2000).
%
\bibitem{gk} W.~T.~Giele and S.~Keller,
Phys.\ Rev.\ D 57, 4433 (1998).
%
\bibitem{zptcdf} T.\ Affolder {\etal.}, (CDF Collaboration),
Phys.\ Rev.\ Lett.\ 84, 845 (2000)
%
\bibitem{d0detector}  S.\ Abachi {\etal.}, (D\O\ Collaboration),
Nucl.\ Instrum.\ Methods Phys.\ Res.\ A 338, 185 (1994).
\bibitem{pdb} D.~Groom {\etal.},
Eur.\ Phys.\ J.\ C 15, 1 (2000).
%
\bibitem{herwig} G. Marchesini {\etal.},
          Comput. Phys. Commun. 67, 465 (1992).
%
\bibitem{lep2}
2001 LEP Electroweak Working Group averages for the Review of Particle Physics,
http://lepewwg.web.cern.ch/LEPEWWG/lepww/4f/PDG01/
%
\bibitem{xbins} G.~Lafferty and T.~Wyatt, 
Nucl.\ Instrum.\ Methods Phys.\ Res.\ A 355, 541 (1995).
%
\bibitem{kstest} This test returns a confidence level $\kappa$ that is
uniformly distributed between 0 and 1 if the two spectra derive from the same 
parent distribution. See, e.g., A.\ G.\ Frodesen, O.\ Skjeggestad, and
H.\ T{\o}fte, ``Probability and Statistics in Particle Physics.'' Columbia
University Press, New York (1979). 
\end{thebibliography}
\end{document}